\documentclass[twocolumn]{aastex631}

\begin{document}

\title{Developing an Automated Detection, Tracking and Analysis Method for Solar Filaments Observed by CHASE via Machine Learning}

\author[0009-0006-5180-8052]{Z. Zheng}
\affiliation{School of Astronomy and Space Science, Nanjing University, Nanjing 210023, China}
\affiliation{Key Laboratory of Modern Astronomy and Astrophysics, Ministry of Education, Nanjing 210023, China}

\author[0000-0002-9264-6698]{Q. Hao}
\affiliation{School of Astronomy and Space Science, Nanjing University, Nanjing 210023, China}
\affiliation{Key Laboratory of Modern Astronomy and Astrophysics, Ministry of Education, Nanjing 210023, China}

\author[0000-0002-1190-0173]{Y. Qiu}
\affiliation{Institute of Science and Technology for Deep Space Exploration, Suzhou Campus, Nanjing University, Suzhou 215163, China}

\author[0000-0002-8002-7785]{J. Hong}
\affiliation{School of Astronomy and Space Science, Nanjing University, Nanjing 210023, China}
\affiliation{Key Laboratory of Modern Astronomy and Astrophysics, Ministry of Education, Nanjing 210023, China}

\author[0000-0001-7693-4908]{C. Li}
\affiliation{School of Astronomy and Space Science, Nanjing University, Nanjing 210023, China}
\affiliation{Key Laboratory of Modern Astronomy and Astrophysics, Ministry of Education, Nanjing 210023, China}
\affiliation{Institute of Science and Technology for Deep Space Exploration, Suzhou Campus, Nanjing University, Suzhou 215163, China}

\author[0000-0002-4978-4972]{M.D. Ding}
\affiliation{School of Astronomy and Space Science, Nanjing University, Nanjing 210023, China}
\affiliation{Key Laboratory of Modern Astronomy and Astrophysics, Ministry of Education, Nanjing 210023, China}

\correspondingauthor{Q. Hao, C. Li}
\email{haoqi@nju.edu.cn, lic@nju.edu.cn}
\begin{abstract}

Studies on the dynamics of solar filaments have significant implications for understanding their formation, evolution, and eruption, which are of great importance for space weather warning and forecasting. The H$\alpha$ Imaging Spectrograph (HIS) onboard the recently launched Chinese H$\alpha$ Solar Explorer (CHASE) can provide full-disk solar H$\alpha$ spectroscopic observations, which bring us an opportunity to systematically explore and analyze the plasma dynamics of filaments. The dramatically increased observation data require automate processing and analysis which are impossible if dealt with manually. In this paper, we utilize the U-Net model to identify filaments and implement the Channel and Spatial Reliability Tracking (CSRT) algorithm for automated filament tracking. In addition, we use the cloud model to invert the line-of-sight velocity of filaments and employ the graph theory algorithm to extract the filament spine, which can advance our understanding of the dynamics of filaments. The favorable test performance confirms the validity of our method, which will be implemented in the following statistical analyses of filament features and dynamics of CHASE/HIS observations. 

\end{abstract}

\keywords{Solar filaments --- Convolutional neural networks --- Astronomy image processing}

\section{Introduction} \label{sec:intro}

Solar filament is one of the typical solar activities in solar atmosphere, which is about 100 times cooler and denser than its surrounding corona \citep{Labrosse2010}. They are observed as dark elongated structures with several barbs, but are seen as bright structures suspended over the solar limb called prominences \citep{Vial2015}. Filaments are always align with photospheric magnetic polarity inversion line (PIL), where the magnetic flux cancellation often takes place \citep{Martin1998,Vial2015}. Filaments sometimes undergo large-scale instabilities, which break their equilibria and lead to eruptions. There is a close relationship among the erupting filaments, flares, and coronal mass ejections, which are different manifestations of one physical process at different evolutionary stages \citep{Gopalswamy2003}. Therefore, the study of the formation, evolution and eruption of filament is not only of great significance to understand the essence physics of solar activities, but also of practical significance for accurately predicting the hazardous space weather \citep{Chen2011,Chen2020}.

Filament are usually observed by ground-based solar H$\alpha$ telescopes around the world, such as Meudon, Big Bear, Kanzelh\"{o}he, Kodaikanal, and Huairou. These telescopes have been the work horses of most of the current knowledge on filaments \citep{Chatzistergos2023}. To study the mechanisms of solar eruptions and the plasma dynamics in the lower atmosphere, the Chinese H$\alpha$ Solar Explorer \citep[CHASE;][]{Li2019,Li2022} was launched into a Sun-synchronous orbit on October 14, 2021. The scientific payload onboard CHASE is the H$\alpha$ Imaging Spectrograph \citep[HIS;][]{Liu2022}, which can provide solar H$\alpha$ spectroscopic observations. It brings us an opportunity to systematically explore and analyze the plasma dynamics of filaments in details. At the same time, the data volume of CHASE/HIS observations has dramatically increased, which also brings challenges to efficiently process such huge amount of data.

In order to statistically obtain the filament features, \citet{Gao2002} developed an automated algorithm combining the intensity threshold and the region growing method. Since then, a number of automated filament detection methods and algorithms based on classical image processing techniques have been developed in the past decades \citep{Shih2003,Fuller2005,Bernasconi2005,Qu2005,Wang2010,Labrosse2010,Yuan2011,Hao2013, Hao2015}. \citet{Shih2003} adopted local thresholds which were chosen by median values of the image intensity to extract filaments. However, this kind of threshold selection cannot guarantee robust results since the bright features on images can significantly affect the value of the thresholds. To overcome this problem, some authors have developed the adaptive threshold methods \citep{Qu2005, Yuan2011, Hao2015}. Particularly, \citet{Qu2005} applied the Support Vector Machine (SVM) technique to distinguish filaments from sunspots. The development of graphics processing unit (GPU) and machine learning in recent years brings a powerful set of techniques which drive innovations in areas of computer vision, natural language processing, healthcare, and also in astronomy in the recent decade \citep{Smith2023, Asensio2023}. Artificial neural networks, especially the convolutional neural networks (CNNs) have been leading the trend of machine learning on the feature segmentation for years since AlexNet \citep{krizhevsky2012}. Recently, CNNs are widely adopt to automated detect filament, and they have been proven to have a high performance \citep{Zhu2019, Liu2021, Guo2022}. \citet{Zhu2019} developed an automated filament detection method based on the U-Net \citep{Ronneberger2015}, a kind of deep learning architecture,which has an excellent performance in semantic segmentation since it can gain strong robustness even from a small data set. Their test performance showed that the new method can segment filaments directly and avoid generating segmentation with a large number of noise points as classical image processing method. A filament may be split into several fragments during its evolution. To figure out whether they belong to one filament is crucial for the study of filament evolution. Many authors adopted morphological operations, the distance criterion and the slopes of the fragments \citep{Shih2003,Fuller2005,Bernasconi2005,Qu2005,Hao2013, Hao2015}. However, these thresholds for distance or angles do not always work. \citet{Guo2022} proposed a new method base on deep-learning instance-segmentation model CondInst \citep{Tian2020,Tian2023} to solve the fragment problems. Since such methods are supervised machine learning methods, we have to first provide data set with filament labeled. The labeled data are usually obtained by manually annotating the ground-based H$\alpha$ images, which cannot maintain the consistence and accuracy of labelling. Thanks to the CHASE mission for providing the seeing-free H$\alpha$ spectroscopic observations, we can get the precise boundaries of filaments for training the deep learning models.

\begin{figure*}[ht!]
    \centering
    \includegraphics[width=\linewidth]{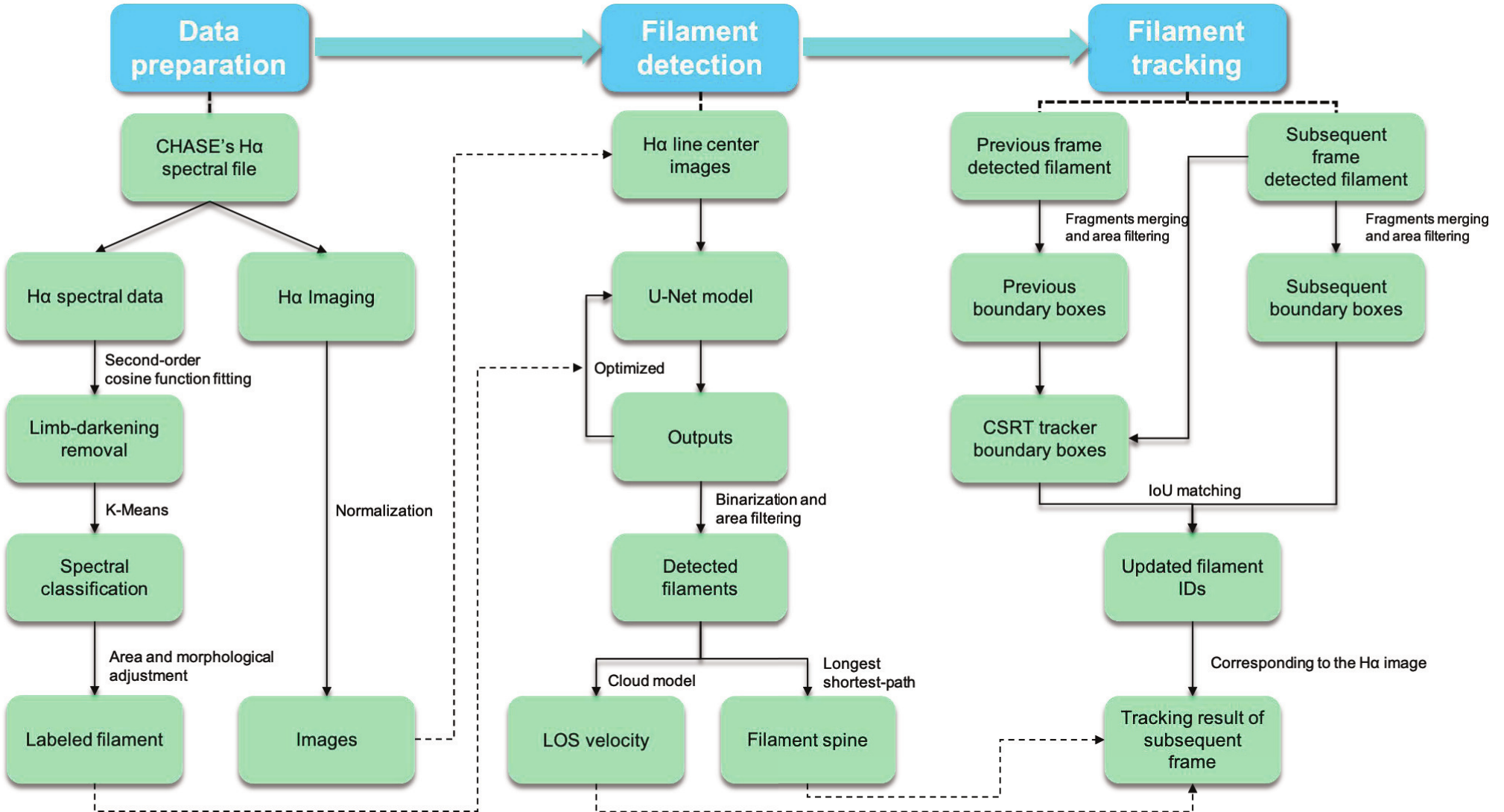}
    \caption{The flowchart of our data preparation, detection, and tracking methods. Solid arrows represent the flow of data processing, while dashed arrows denote data connections between different stages of the processing.
    \label{fig1}}
\end{figure*}

In this paper, we developed an efficient and robust automated detection and tracking method for filament observed by CHASE. Figure~\ref{fig1} shows the flowchart of our method consisting of three parts: data preparation, filament detection, and filament tracking. In Section \ref{data_preparation}, we describe the data preparation, including the filament labeling, calibration for image data, and other necessary adjustments. The pipelines for the automated detection and tracking system are descried in Section~\ref{detection} and ~\ref{tracking}, followed by the description of performance, respectively. Section~\ref{feature_extraction} is dedicated to the inversion of line-of-sight velocity and filament spine extraction. Discussion and conclusions are given in Section~\ref{discussion&conclusion}.

\section{Data preparation} \label{data_preparation}

\begin{figure}[ht!]
    \centering
    \includegraphics[width=1.\linewidth]{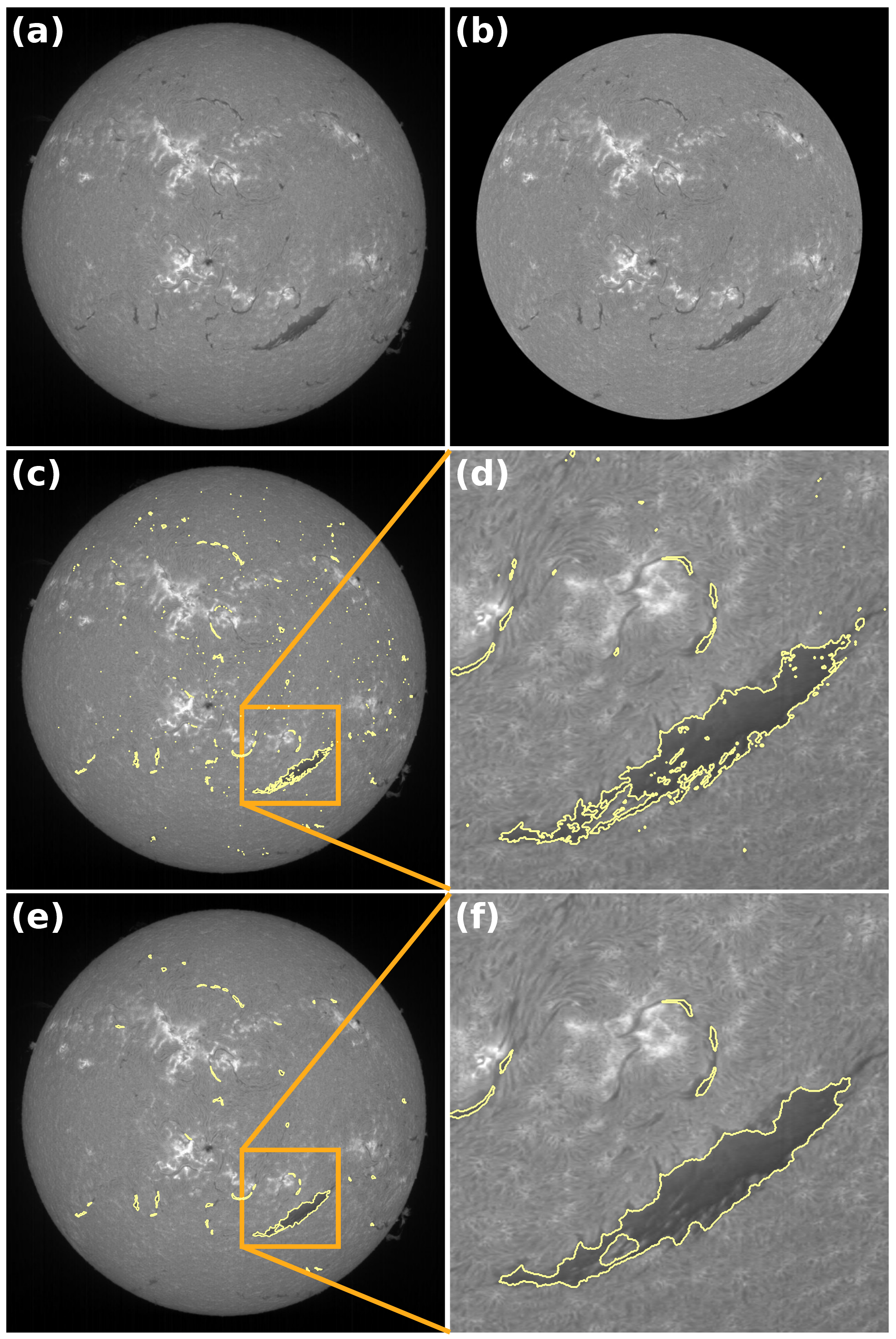}
    \caption{An observation obtained on 2023 January 19 as an example to show the schematic of data preparation. (a) The original H$\alpha$ line center image. (b) The image after limb-darkening removal. (c)  The K-Means spectral classification of filament candidates with yellow contours. (d) An enlarged part of (c). (e) and (f) is similar to (c) and (d), but is the final labeled filament after area filtering and morphological close operation. 
    \label{fig2}}
\end{figure}

The CHASE/HIS implements raster scanning of the full solar disk within 60 seconds at the wavebands of H$\alpha$ and Fe I with a spectral sampling of 0.048~\AA \ and a pixel resolution of $1.04^{\prime\prime}$ in the binning mode \citep{Qiu2022}. Although our detection and tracking method are based on the H$\alpha$ line center images, the high quality H$\alpha$ spectral information allows us to get the precise boundaries of filaments by spectral classification for training the deep learning models. As shown in Figure~\ref{fig1}, before the filament detection and tracking, we proposed a series of preprocessings to build a data set for training and testing the deep-learning model. First, we used a second-order cosine function to remove the limb darkening. Next, we used a unsupervised learning method called K-means \citep{MacQueen1967} to classify the whole disk spectra in order to get the precise boundaries of filaments. The following subsections explain the details of each approach.

\subsection{Spectral Classification for Filament Labeling} \label{spectral_classification}

Our filament detection method is a supervised machine learning method, which means we need to provide the data sets with filament labeled. Here we utilize the CHASE full-disk H$\alpha$ spectra to identify and label the filament. It is difficult to identify filaments by spectra manually since there are 118 wavelength points in CHASE H$\alpha$ profile. We employed K-Means algorithm \citep{MacQueen1967}, an unsupervised clustering algorithm which has been widely applied in spectral classification of solar physics \citep{Viticchie2011,Panos2018,Asensio2023}.  It requires the initial setting of the number of clusters, denoted as $k$. Subsequently, it will assign $k$ centroids $(\mu_1,\mu_2,\cdots,\mu_k)$ and divide the data into $k$ clusters $(S_1,S_2,\cdots,S_k)$ based on the distance to the centroids. Then, it continually updates the centroids and clusters to minimize the intracluster distance by solving following equation:
\begin{equation}
    \mathop{\arg\min}_{\mu} \sum_{i=1}^{k} \sum_{x_j \in S_i}
    \Vert x_j - \mu_i \Vert,
\end{equation}
where $x$ is the contrast of the spectral intensities of each pixel with the average spectral intensity $I/I_{avg}$.

The K-means method is very sensitive to the uneven radiation intensity distribution across the solar disk, which is adversely affecting spectral classification for labeling filaments. We use a second-order cosine function \citep{Pierce1977} to remove the limb darkening:
\begin{equation}
    I_{\lambda}^{*}(R) = a_{\lambda} + b_{\lambda} * cos\theta + c_{\lambda} * cos^{2}\theta,
\end{equation}
where $I_{\lambda}^{*}(R)$ is the mean observed radiation intensity at the radius $R$ in wavelength $\lambda$, $R_S$ is the radius of the Sun, $cos\theta =\sqrt{1-(R/R_s)^2}$ and $a_{\lambda}$, $b_{\lambda}$, $c_{\lambda}$ are the parameters to be fitted. We used the radiation intensity along the solar equator as the fitting data and applied a least-squares fit to minimize the deviation rate. The radiation intensity along the solar equator is selected since there are few activities and it is longer enough to fit the whole disk. For each wavelength $\lambda$ of the H$\alpha$ profile, we need find the best $a_{\lambda}$, $b_{\lambda}$, $c_{\lambda}$ to minimize the following equation:
\begin{equation}
    \sum_{p_i \in eq} \mid I_{\lambda}(p_i) - I_{\lambda}^*(R(p_i)) \mid / I_{\lambda}^*(R(p_i)),
\end{equation}
where $I_{\lambda}^*(p_i)$ is the observed radiation intensity at point $p_i$ of the equator and $R(p_i)$ is the radius at $p_i$. We limited the fitting to 880 pixels (about $0.95R_s$) since the pixels alone the edge may vary with the observation. Figure~\ref{fig2}(a) and (b) give an example to show the H$\alpha$ line center image before and after limb-darkening removal, respectively.

After a series of trial and error we set the $k=30$, i.e., the K-Means algorithm automated categorizes the spectra into 30 classes, in which they belong to sunspot, plage, filament and so on. Then we manually select the class that most closely matches the filament as the output label candidate. As shown in Figure~\ref{fig2}(c) and (d), the K-Means spectral classification could effectively segment filaments from the solar disk. However, small chromospheric fibers can not be removed since they are also cold material and have similar H$\alpha$ spectral profiles to the filaments. Here, we set a area threshold of 64 pixels ( about 69 arcsec$^2$) to sieve the chromospheric fibers. We can also find that the filament classified by K-Means are more mottled with holes as shown in Figure~\ref{fig2}(d). We adopt the morphological close operation with round structure element (radius is 5 pixel) to fill these holes. Figure~\ref{fig2}(e) and (f) show the results after adjustment. Then the labeled data are used for model training. In addition, we found that the data with morphological close operation can improve the ability of U-Net model to distinguish the sunspot and filament. It is because the input of U-Net model is the normalized H$\alpha$ line center images, which can not distinguish the sunspot and filament only by intensity since they have similar dark appearance.

\subsection{Data arrangement for model training} \label{other_adjustments}

We collected 120 sets of H$\alpha$ spectral observation from CHASE and constructed our labeled dataset by applying the approaches mentioned above. The data span from December 2022 to July 2023, with a time interval of about 2 days. In a supervised machine learning method, the data usually are divided into training, validation, and testing set, respectively. The validation set was utilized during the training phase to choose models, which can prevent overfitting on the training set. The testing set was employed to assess the performance of trained models. Therefore, these 120 sets are divided into 20 groups, the third and the sixth sets in each group are chosen as the validation and testing sets, respectively, while the others are chosen as the training set, with the ratio being about $1:1:4$. In this way, we can ensured a evenly distribution in our data sets so that the trained model could be robust and increase its generalization performance. 

The data augmentation technique is usually applied to enlarge the dataset by flipping and rotating transformations since the convolution kernel does not have rotational symmetry and axis-symmetry, which can reduce the risk of overfitting \citep{Miko2018}. Here, we also adopt data augmentation during the training process to enhance the robustness of our model. In each training iteration, the data are randomly applied transformations, including vertical flipping (with a $50\%$ probability) and rotation with a angle within $[-45^{\circ}, 45^{\circ}]$ (with an $80\%$ probability).

\begin{figure*}[ht!]
    \centering
    \includegraphics[width=1.\linewidth]{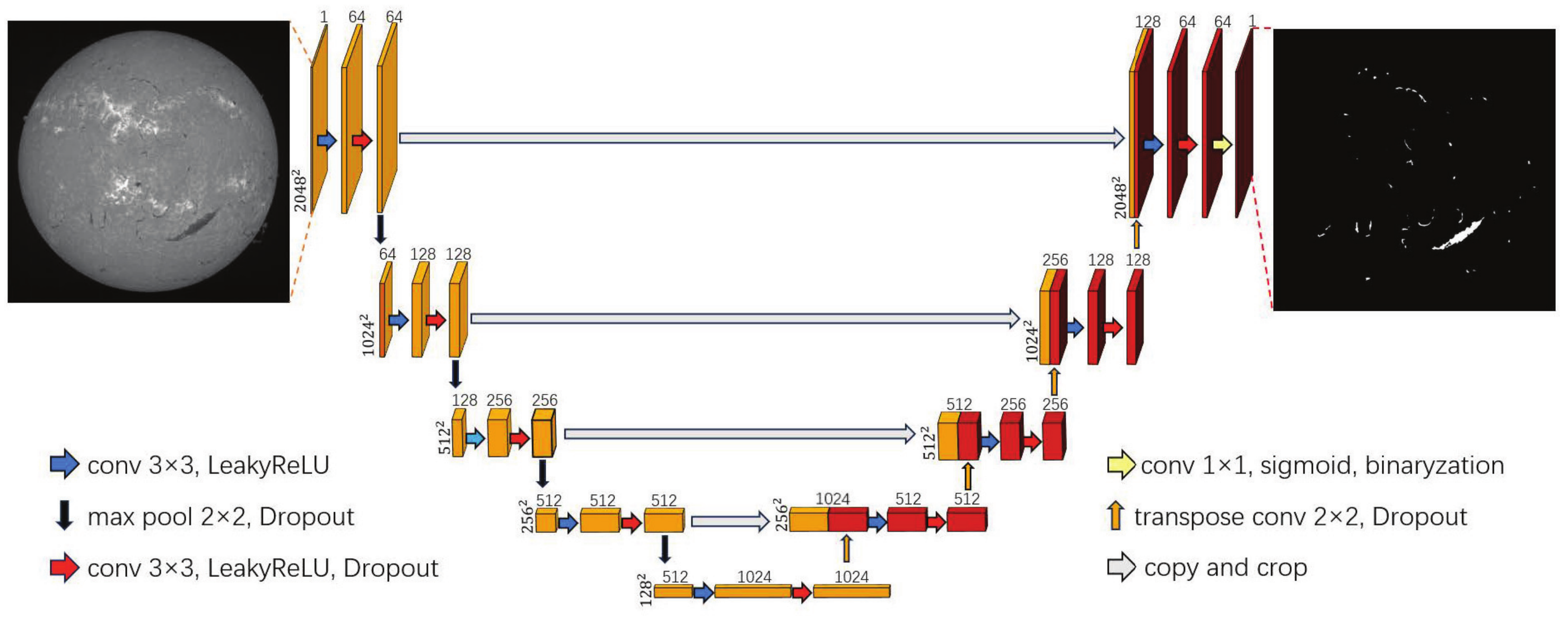}
    \caption{Schematic representation of our U-Net model. The model takes H$\alpha$ line-center images as input and produces a binary image of the same size as output. The cubes represent feature maps, where the dimension of each map is indicated on its left, and the number of channels is indicated above it. Operations for each channel are represented by arrows with different colors.
    \label{fig3}}
\end{figure*}

\section{Filament detection} \label{detection}

Recently, CNNs are adopt to automated detect filament, and it have been proven to have a high performance \citep{Zhu2019, Liu2021, Guo2022, Zhang2023}. Here we employ the U-Net \citep{Ronneberger2015} architecture, a kind of CNN architecture, to implement filament detection in our work. Figure~\ref{fig1} gives the flowchart of the detection method. Compared to using H$\alpha$ spectral data as the input of  K-means method during the preprocessing steps, we only adopt the H$\alpha$ line center image as the input of U-Net model. Note that the input images are normalized by dividing its mean intensity instead of the maximum intensity since the errors arising from variations in observation times, especially during the flaring time.

\subsection{U-Net architecture for filament detection} \label{U-Net}

The U-Net model derives its name from its U-shaped architecture, as shown in Figure~\ref{fig3}. Upon receiving image inputs, it initiates an encoding process to progressively extract high-level semantic information while reducing the image dimensions, as shown in left part of Figure~\ref{fig3}. Following this, the decoding process reconstructs the feature maps to the original image size (right part of the U-shaped architecture), delivering detailed pixel-to-pixel level prediction results.

In our model, the encoding and decoding parts each has four blocks, which are repeating patterns of layers. During the encoding process, each block has two $3\times3$ convolution layers, followed by a $2\times2$ max-pooling layer, as indicated by the blue, red, and black arrows in the left side of Figure~\ref{fig3}. A convolution layer works as a feature detector to get the feature maps. As the layers of the neural network go deeper, the number of the output channels gets larger, which can regard as the feature numbers increasing. Max-pooling layers output a feature map containing the most prominent features of the previous feature map, which play a critical role in compressing the size of feature maps while preserving important features and relationships in the input images. For example, the input is an image with 1 channel and dimension size of $2048 \times 2048$, the output after the first block becomes a feature map with 64 channels and dimension size of $1024 \times 1024$, as shown in Figure~\ref{fig3}. The feature maps from the convolution layer passes on to activation functions to make the network adapt to nonlinear features from the previous layers. We use LeakyReLU and Sigmoid functions as our activation functions, which are defined as:

\begin{equation}
    LeakyReLU_{\alpha}(x)=\left\{
    \begin{array}{cc}
    \alpha x, & {x < 0} \\
    x, & {x \geq 0}
    \end{array};\right.
\end{equation}

\begin{equation}
    Sigmoid(x)=1/(1+e^{-x}),
\end{equation}

where $\alpha$ is a hyperparameter and set to $0.01$ in our model. The Sigmoid function is only applied in the output layer of the model to transform the output into the range between 0 and 1 for binary classification, enabling the distinction between targets, i.e., filaments and the background.

In the decoding process, the feature and spatial information through a series of transpose convolutions return to the original image dimension. Transpose convolution, often referred to as deconvolution or up-sampling, involves the introduction of zero values through interpolation in the image, followed by a convolution operation, allowing for effective image magnification. After this block, the output feature map size are 4 times larger, from $128 \times 128$ to $256 \times 256$, $256 \times 256$ to $512 \times 512$ and so on, as indicated by Figure~\ref{fig3}. Copy and crop, or skip connections, indicated by the gray arrows in Figure~\ref{fig3}, entails the concatenation of semantic information derived from the encoding process with the corresponding feature maps during the decoding process. This operation facilitates the transmission of augmented semantic information, thereby bolstering the segmentation performance of the model. 

Loss functions play an important role in machine learning. They define an objective by which the performance of the model is evaluated. The parameters learned by the model are determined by minimizing a chosen loss function. In other words, loss functions are a measurement of how good our model is at segmenting the filaments. We utilize the Focal Loss function \citep{Lin2017}, a well-established choice in binary classification scenarios:

\begin{eqnarray}
	FocalLoss_{\gamma, w}(y,\hat{y})=\sum- w y (1-\hat{y})^{\gamma} \ln \hat{y}  \nonumber\\
	- (1-y) \hat{y}^{\gamma} \ln (1-\hat{y}),
\end{eqnarray}

where $y$ represents the values in the label, and $\hat{y}$ is the probability values output by the model. $\gamma$ is a hyperparameter for reducing the relative loss for well-classified examples, putting more focus on hard, misclassified examples, i.e., filament regions. $w$ stands for the weight of filament regions. Focal Loss proves beneficial in addressing the challenges posed by imbalanced distribution of positive and negative samples, corresponding to the scenario where the filament regions are much smaller than the non-filament regions. The training of the U-Net model involves the iterative adjustment of the weights of various convolution kernels within the convolution layers to minimize the Focal Loss by the optimizer, which we choose ADAM optimizer \citep{Kingma2014} in our model. A dropout layer is a regularization technique employed during model training to stochastically zero out a subset of weights, which can mitigate overfitting in the model. We also adapt dropout layers in our model. The detailed parameter settings are list in Table~\ref{tab1}.

\begin{table}[!ht]
    \centering
    \caption{Parameters of Our U-Net Model.}
    \label{tab1}
    \begin{tabular}{l|c}
    \hline
        Convolution Kernels & See Figure~\ref{fig3} \\ \hline
        Active functions & See Figure~\ref{fig3} \\ \hline
        Optimizer & ADAM \\ \hline
        Epochs & 500 \\ \hline
        Batchsize & 1 \\ \hline
        Learning rate & 1.0$\times 10^{-5}$ \\ \hline
        $\alpha$ in LeakyReLU & 0.1 \\ \hline
        Dropout rate & 0.2 \\ \hline
        $w$ in Focal Loss & 2 \\ \hline
        $\gamma$ in Focal Loss & 4 \\ \hline
    \end{tabular}
\end{table}

\subsection{Performance of detection} \label{performance_detection}
The precision $P$ and recall ratio $R$ are common semantic segmentation evaluation strategy which we also adopt in our model evaluation. These two metrics are defined as:
\begin{eqnarray}
	P = \frac{TP}{TP + FP}\,,\\
	R = \frac{TP}{TP + FN}\,,
\end{eqnarray}
where TP, FP, and FN denote the true positive, false positive, and false negative measurements, respectively. In addition, the intersection over union (IoU) is often used to evaluate the performance of a model \citep{Rezatofighi2019}, which is defined as:
\begin{equation}
	\text{IoU} = \frac{TP}{TP + FP + FN} = \frac{R*P}{R+P-R*P}\,\,.
\end{equation}
In practical, the IoU over $60\%$ is a good enough score. The average precision and recall ratios of the training, validation, and testing are listed in Table~\ref{tab2}. The precision, recall, and IoU ratios of each data file are also plotted in Figure~\ref{fig5}. These results indicated our model is a viable strategy for solar filament recognition.

\begin{table}[!ht]
    \centering
    \caption{Performance of Our U-Net Model.}
    \label{tab2}
    \begin{tabular}{c|c|c|c}
    \hline
          & IoU & Precision & Recall \\ \hline
        Training set & 0.69 & 0.79 & 0.85 \\ \hline
        Validation set & 0.67 & 0.78 & 0.84 \\ \hline
        Testing set & 0.67 & 0.79 & 0.83 \\ \hline
    \end{tabular}
\end{table}

\begin{figure}[ht!]
    \centering
    \includegraphics[width=1.\linewidth]{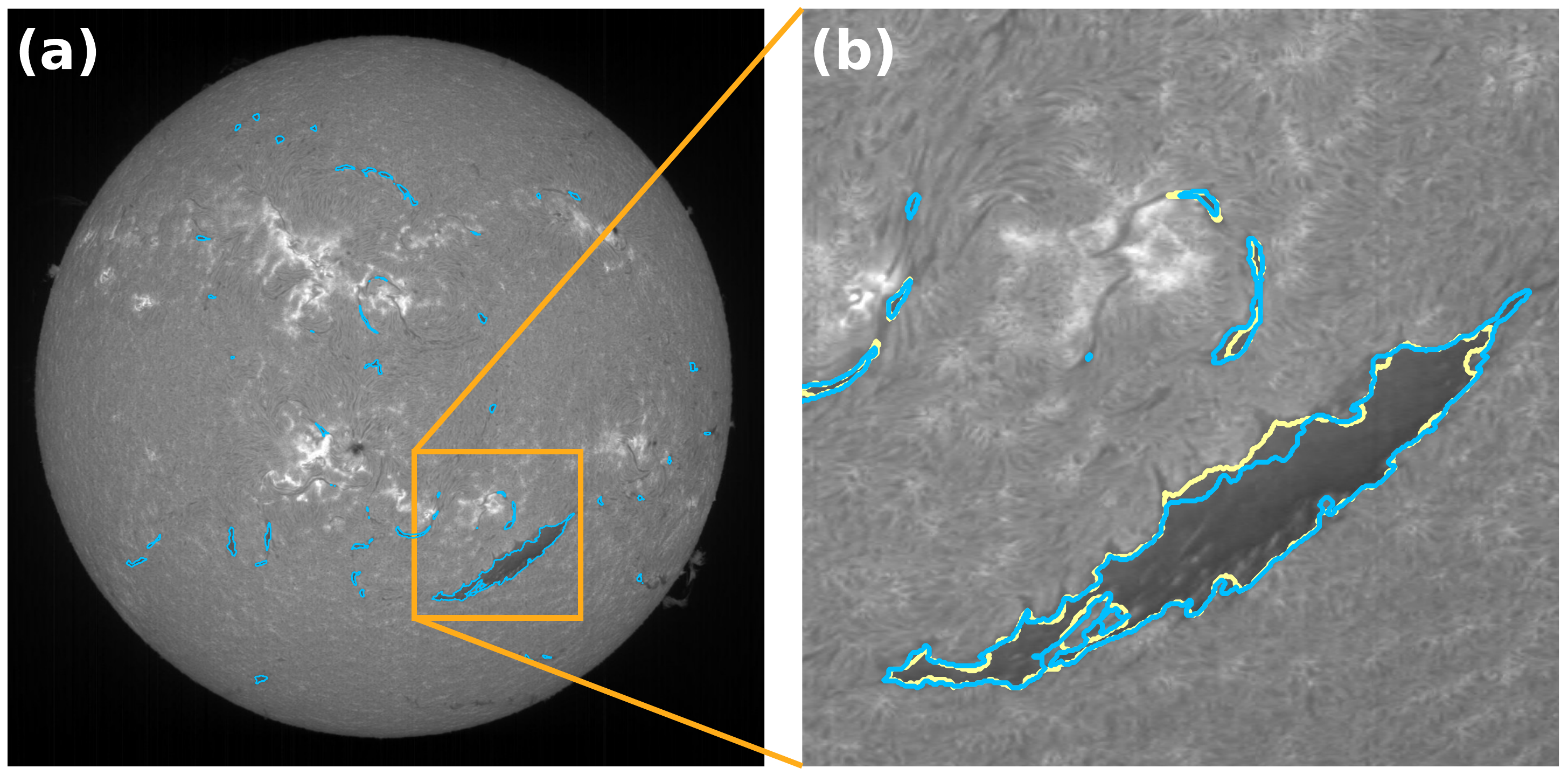}
    \caption{
    The detected results by our U-Net model. (a) The detected filaments with blue contours are plotted above the original H$\alpha$ line center images. (b) A enlarged part of (a), where the ground-truth indicated by the yellow contours. }
    \label{fig4}
\end{figure}

\begin{figure}[ht!]
    \centering
    \includegraphics[width=1.\linewidth]{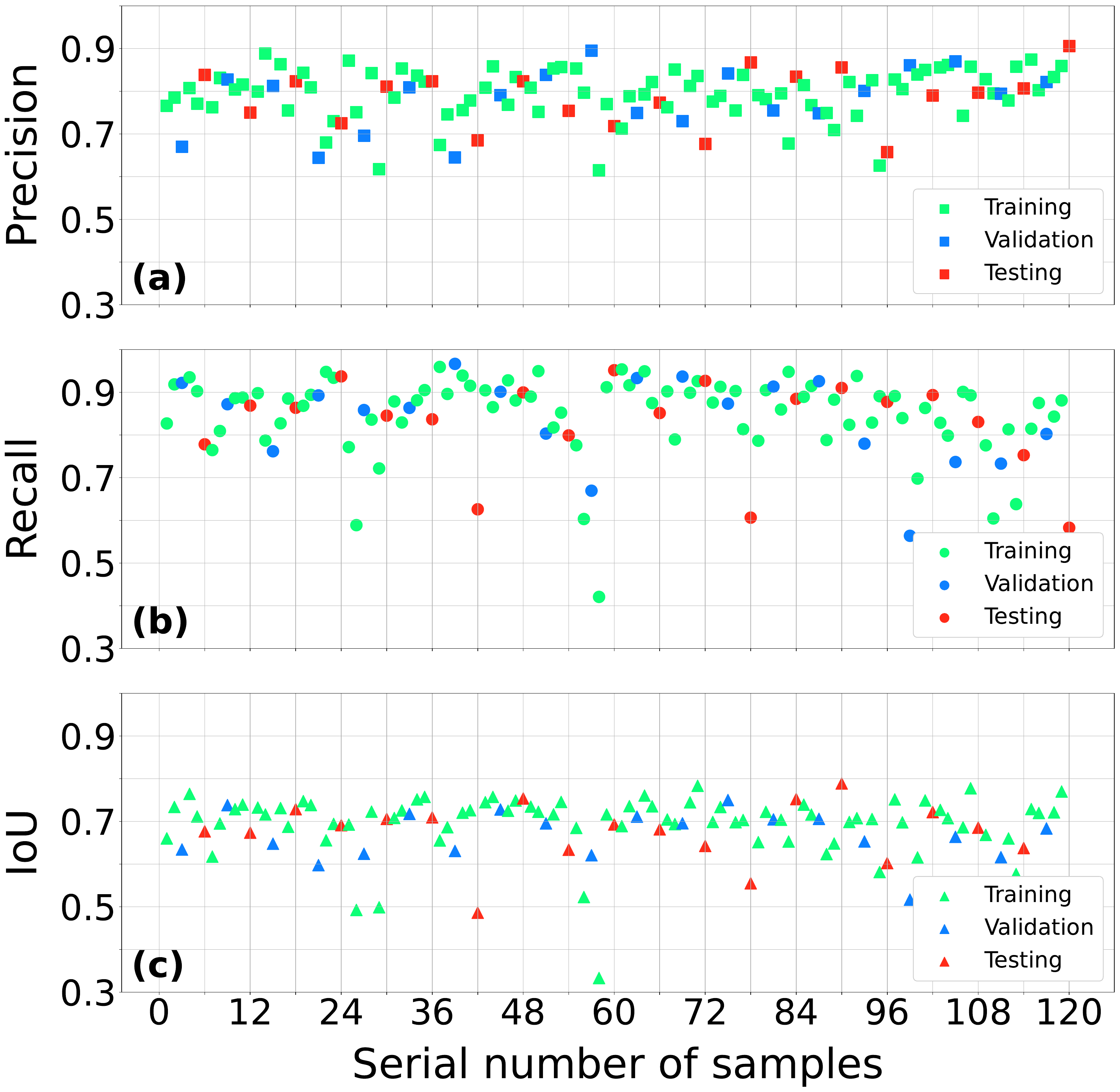}
    \caption{
    The performance of our U-Net model. (a), (b), and (c) show the precision, recall, and IoU ratios of the samples in the training, validation, and testing set, respectively.}
    \label{fig5}
\end{figure}

\begin{figure*}[ht!]
    \centering
    \includegraphics[width=1.\linewidth]{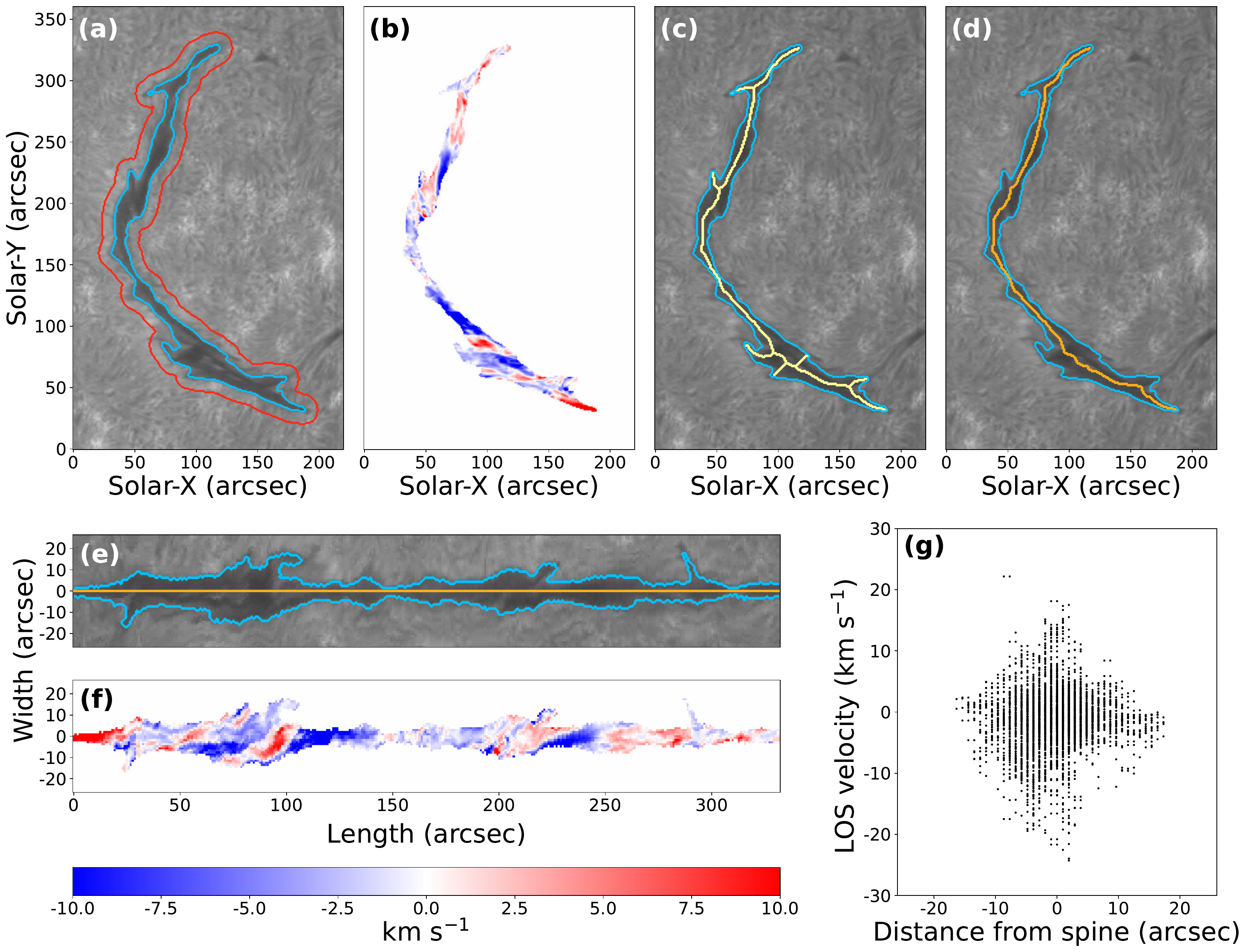}
    \caption{
    A detected filament of the CHASE observation obtained on 2023 April 25 as an example to show the schematic of filament feature extraction. (a) The detected filament with blue contour, between the red and blue contours are the region set as the quiet region for LOS velocity inversion. (b) The LOS velocity distribution of the filament. (c) The filament skeleton marked with yellow curve. (d) The filament spine marked with orange curve. (e) The straightened filament. (f) The corresponding LOS velocity distribution of (e). (g) The scatter plot of LOS velocity perpendicular to the filament spine.}
    \label{fig6}
\end{figure*}

\section{Filament Feature Extraction} \label{feature_extraction}

After the filament detection, we can obtain the basic morphological features such as filament location and area. These morphological features are not enough for the analysis of the evolution and eruption of filaments, so we apply the cloud model to invert line-of-sight (LOS) velocities of filaments and employed the graph theory algorithm to extract the filament spines, which are explained in following two subsections.

\subsection{LOS velocity inversion of filaments} \label{LOS_velocity}

The dynamic evolution of filaments has consistently been a key issue in the study of filament \citep{Chen2020}. CHASE provides the full-disk H$\alpha$ spectral data, which bring us a chance to extract the dynamic information of filaments. We applied the cloud model \citep{Beckers1964} to fit the H$\alpha$ spectra of filaments in order to derive its LOS velocities. The contrast function of cloud model refers to:

\begin{equation}
    C(\lambda) = \frac{I_0(\lambda) - I(\lambda)}{I_0(\lambda)} = 
    (1 - \frac{S}{I_0(\lambda)})(1-e^{(-\tau)}),
\end{equation}

where $\tau = \tau_0 e^{-(\lambda - \lambda_0 - v\lambda_0/c)^2/w^2}$, $I$ represents the spectral profile of the filament region,  $I_0$ corresponds to the quiet region, $\lambda_0$ is the line center wavelength, and $c$ is the light speed. The parameters to be fitted include the LOS velocity of the cloud $v$, optical depth $\tau_0$, source function $S$, and Doppler width $w$.

The cloud model inversion is applied to derive the spectral profiles of each individual filament. As shown in Figure~\ref{fig6}(a), the blue contour represents the filament detected by our U-Net model. The red contour represents the extension of the blue one by 10 pixels. The average spectrum between the red and green contours is extracted as $I_0$. we determine the line center wavelength $\lambda_0$ by using the moment method \citep{Yu2020}. Then, the contrast profile of the spectrum in blue contour is applied by a least-squares fitting, providing the inversion result of the spectral profile. Figure~\ref{fig6}(b) presents the inversion result of LOS velocity distribution of the entire filament region.

\subsection{Filament spine extraction} \label{filament_spine}

Filament spine defines the skeleton of a filament along its full length, with several extend barbs. In order to derive the filament spine, many authors employed iterations of the morphological thinning and spur removal operations \citep{Fuller2005, Qu2005, Hao2013}. However, after iterated spur removal operations the spines often become shorter than the original ones. \citet{Bernasconi2005} used the Euclidean distance method to overcome the shortcoming by morphological spur operation. \citet{Yuan2011} and \citet{Hao2015,Hao2016} used the algorithm based on the graph theory by calculating the paths from end to end points of filament skeleton in pixels, where the longest path is kept as the spine. This method is also employed in our work.

Following the acquisition of the LOS velocity field distribution of filaments, we want to systematically analyze the evaluations of various filaments. However, different filaments have different length, we need compare them in a standardized manner. Thus, we straighten and normalize the filaments along their spines. This approach can help to uncover unified patterns in the dynamical evolution of various kind of filaments. After extracting the spine of the filament, we could obtain the distribution of the coordinates $\{(x_i,y_i)\}_{i=1}^{n}$ of points along the spine with respect to the distance set $\{l_i\}_{i=1}^{n}$, where $n$ is the number of pixels along the spine. Subsequently, we perform spline interpolation then obtain the parametric equation of the main spine:

\begin{equation}
    \left\{\begin{array}{cc}
    x = X(l), \\
    y = Y(l).
    \end{array}\right.
\end{equation}

Then, we differentiate the parametric equations and apply Gaussian filter (with $\sigma=5$) for smoothing.  With the tangent equation $(x^{\prime},y^{\prime})=(\mathrm{d}X/\mathrm{d}l,\mathrm{d}Y/\mathrm{d}l)$, we can obtain the corresponding normal equation $(y^{\prime}, -x^{\prime})$. Subsequently, we shift each point on the major axis in the direction perpendicular to the normal by the size of $2S_{f}/L_{f}$, where $S_{f}$ is the filament area and $L_{f}$ is the spine length, respectively. This process can straighten filaments along their spines according to their irregular shapes. Figure~\ref{fig6} gives an example of our process. The yellow and orange curves shows the derived filament skeleton and spine in Figure~\ref{fig6}(c) and (d), respectively. Figure~\ref{fig6}(e) shows the straightened filament, which also effectively preserves the morphological characteristics of filaments, such as brab structures.

Moreover, the straightening approach enables a quantitative study of the distribution of physical information along and perpendicular to the spine. As shown in Figure~\ref{fig6}(f) and (g), the LOS velocity distribution along and perpendicular to its spine. Note that the majority of filaments exhibit a rhombic or elliptical LOS velocity distribution, suggesting that the locations with the maximum LOS velocity in filaments are typically near the spine.

\begin{figure*}[ht!]
    \centering
    \includegraphics[width=1.\linewidth]{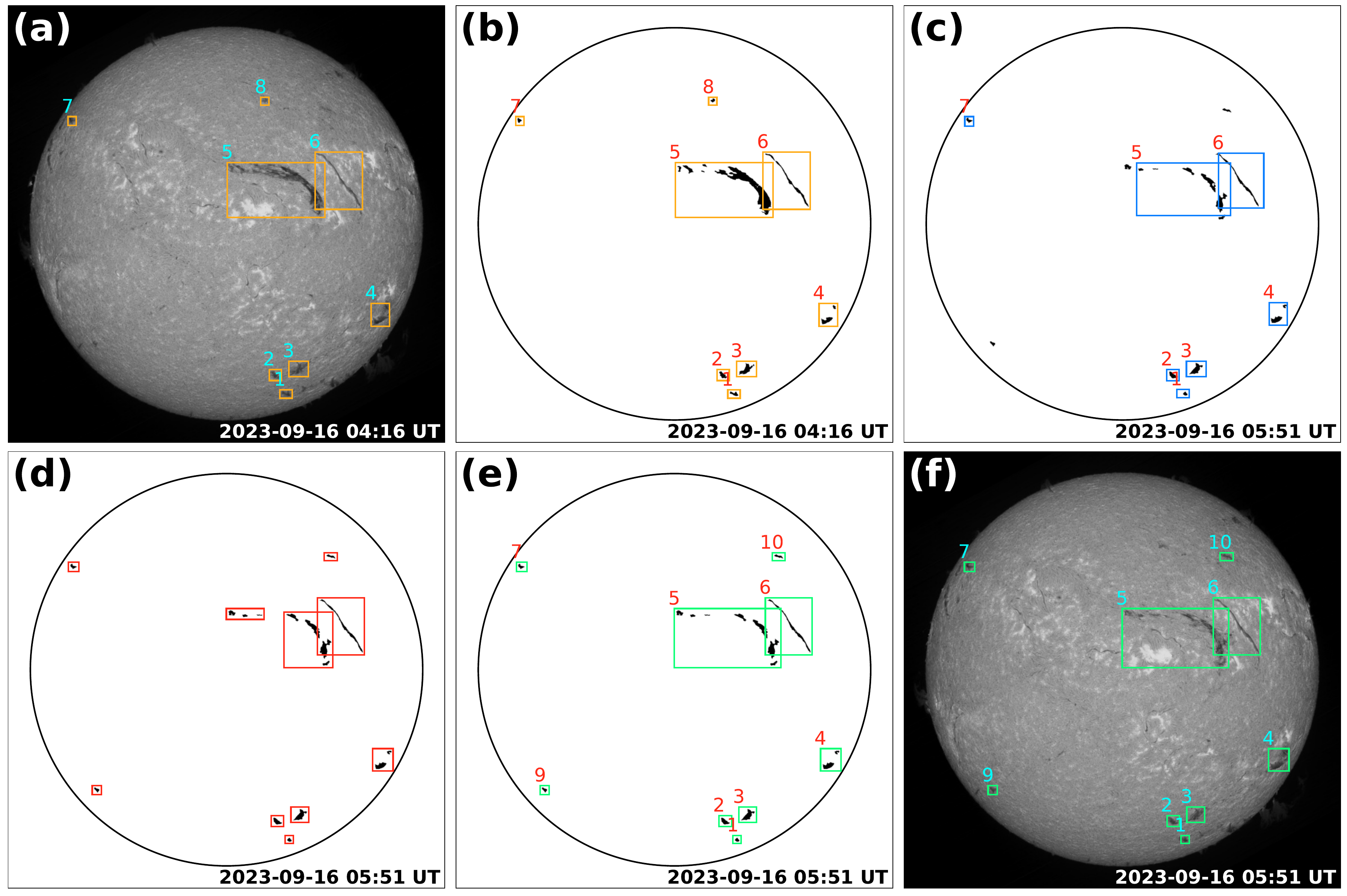}
    \caption{An example for explaining our filament tracking method. (a) The previous frame H$\alpha$ line center images observed at 2023 September 16 04:16 UT. The orange, blue, and yellow boxes with ID numbers indicate the tracking results in different steps. (b) Similar to (a), but without the background image. The black regions are the detected filaments by our U-Net model. (c) The tracking results of the subsequent frame observed at 2023 September 16 05:51 UT by the CSRT tracker. (d) The results of the subsequent frame after the initialized step. Each red box represent a merged filament. (e) The final results after update step of the subsequent frame. (f) Similar to (e), but with H$\alpha$ line center images as its background.}
    \label{fig7}
\end{figure*}

\begin{figure*}[ht!]
    \centering
    \includegraphics[width=\linewidth]{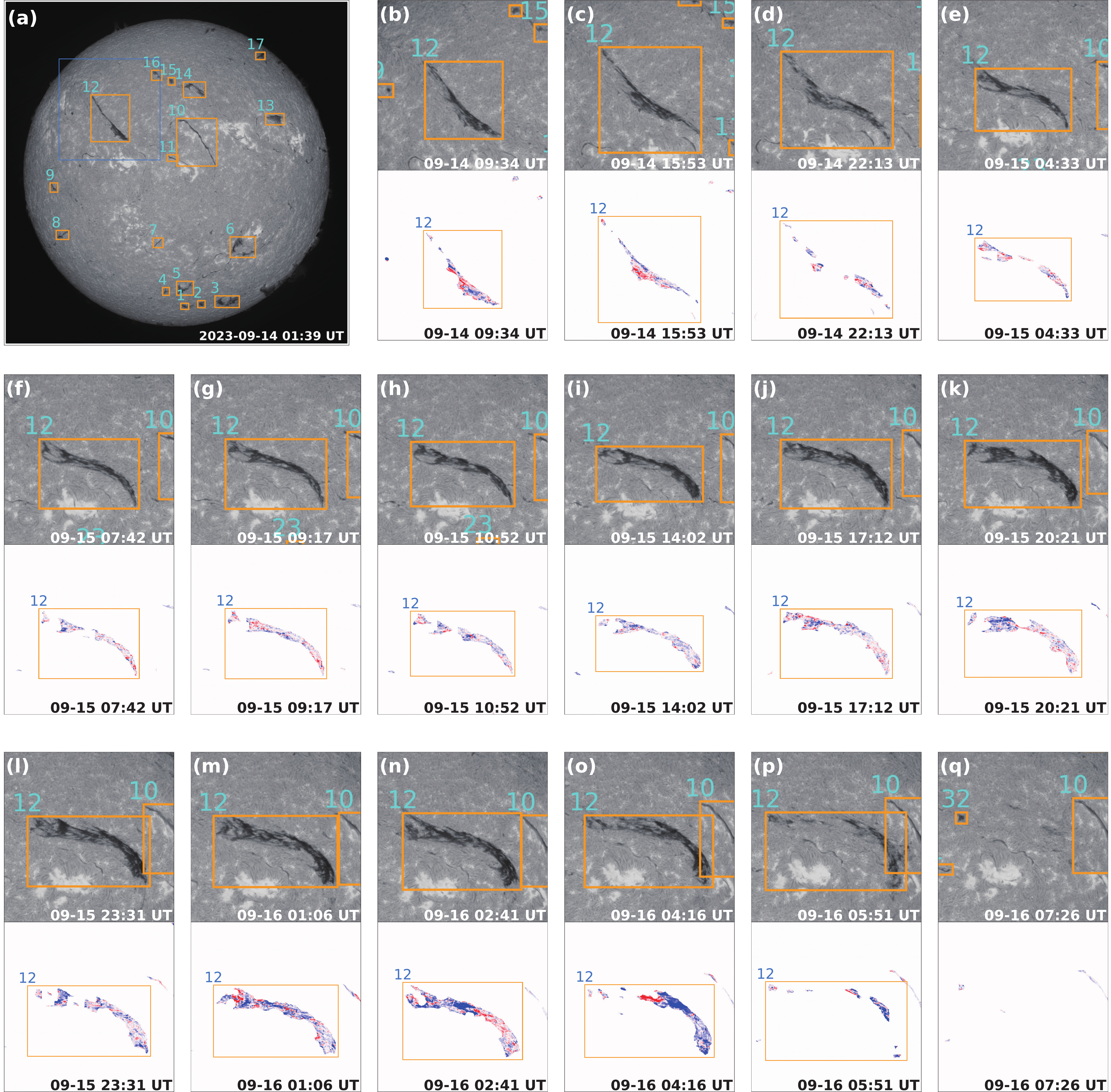}
    \caption{
    An example showing the tracking results. (a) The first frame of H$\alpha$ line center image. The orange boxes with numbers indicate the tracking results. (b--p) A series of filament tracking results. Each panel has the same field-of-view as the blue box in (a), and the top sub-panel is the H$\alpha$ line center image and the bottom one is the corresponding LOS velocity distribution.}
    \label{fig8}
\end{figure*}

\section{Filament Tracking} \label{tracking}

Automated filament tracking assumes a pivotal role in statistical analysis of filament evolution. \citet{Hao2013} considered the detected filament locations, areas, and the differential rotations to trace the daily filament evolution. \citet{Bonnin2013} first retrieved the location and morphology of filament main skeleton by the filament automated recognition method by \citet{Fuller2005}, then applied a curve-matching algorithm to determine the filaments in different frames whether they are the same one. Actually, filament sometimes split into several fragments or partially erupted, which represent intricate and dynamic evolution \citep{Shen2012,Liu2012,Sun2023,Hou2023}. Tracking filament by extracting and parameterizing certain morphological features, is only effective for relatively larger filament and shorter time intervals. Here, we propose the Channel and Spatial Reliability Tracking (CSRT) algorithm without requiring additional feature extraction of transformation manually. The CSRT method is proficient in tracking moving and deformable targets, which is quite suitable for filament tracking.

\subsection{Tracking method}

The CSRT tracker is C++ implementation of the CSR-DCF (Channel and Spatial Reliability of Discriminative Correlation Filter) tracking algorithm \citep{Lukezic2017} in OpenCV library \citep{opencv}. The tracked object is localized by summing the responses of the learned correlation filters and weighted by the estimated channel reliability scores.  In other words, the CSRT tracker distinguish the target and background based on adjusting the weights of different channels, e.g., 10 HoGs (histogram of oriented gradients) channels and intensity channel for grayscale-images. After that, the target is localized by the probability map and its region is updated. Furthermore, \citet{Farhodov2019} integrated the region-based CNN pre-trained object detection model and the CSRT tracker, and got better tracking results since the detection model has already separated the traget and the background. Therefore, we integrate the U-Net model and the CSRT algorithm, i.e., the filaments detected by the U-Net model are used as inputs of the CSRT tracker.

The third part of Figure~\ref{fig1} shows our tracking processing scheme, which consists of two part, i.e., the initialization step and the update step. During the initialization step, we need to set a series of boxes which contain the detected filaments in each frame since the input of the CSRT tracker is an image with targets with bounding boxes. A simple way is using the minimum boundary boxes of the detected filaments directly as the input. However, filament sometimes may split as several fragments, or several fragments may merge into one filament. We adopt a distance criterion of 50 pixels (about 52$^{\prime\prime}$) to combine filament fragments and a area criterion of 200 pixels (about 216 arcsec$^{2}$) to filter relative small filament fragments. These operation can enhance the stability and accuracy of the CSRT tracker. Figure~\ref{fig7} shows two frames of H$\alpha$ line center images observed at 04:16 UT and 05:51 UT on 2023 September 16 as the previous and subsequent frames, respectively. The orange boxes in Figure~\ref{fig7}(a) and (b) are the results after the initialization step, where each box indicates the merged single filament. The previous frame has been marked with a unique tracking ID, as shown by the numbers above the boxes in Figure~\ref{fig7}(a) and (b). Then the CSRT tracker tracks the subsequent frame and output the tracked filaments based on the previous tracking IDs. Figure~\ref{fig7}(c) shows the tracked filament indicated by the blue boxes, which have the same tracking IDs as that in previous frame in Figure~\ref{fig7}(b). 

However, in addition to the fragmentation of filaments, it is possible that the filament maybe disappear after eruption as well as the formation of a new filament. Therefore, in the update step, we adopt the IoU ratio to compare the tracking result of CSRT tracker with the results of the subsequent frame after initialization step. If the filament (merged boundary boxes) of the subsequent frame after initialization step has the largest IoU with a certain tracked filament by the CSRT tracker, it will be set the same tracking ID; if the IoU is empty, i.e., there is no corresponding filament, it will be set a new ID. In this way, the update step is finished and the result will be input to CSRT tracker again for tracking the next subsequent frame until all frames are tracked. The red boxes in Figure~\ref{fig7}(d) are the results after the initialization step, which has no IDs yet. After the update step, the red boxes in Figure~\ref{fig7}(d) and blue boxes Figure~\ref{fig7}(c) are compared by their IoU ratios and finally output the tracking results indicated by the yellow boxes in Figure~\ref{fig7}(e) and (f). The update step can effectively handle with the situation of the eruption of a filament. We extracted a filament with ID  No.12 within field-of-view of the blue box in Figure~\ref{fig8}(a) as an example. We can see its dynamic evolution and eruption from Figure~\ref{fig8}(b) to (p) at different time, and finally disappeared after eruption in Figure~\ref{fig8}(q).

\subsection{Performance of tracking} \label{performance_tracking}

\begin{figure}[ht!]
    \centering
    \includegraphics[width=\linewidth]{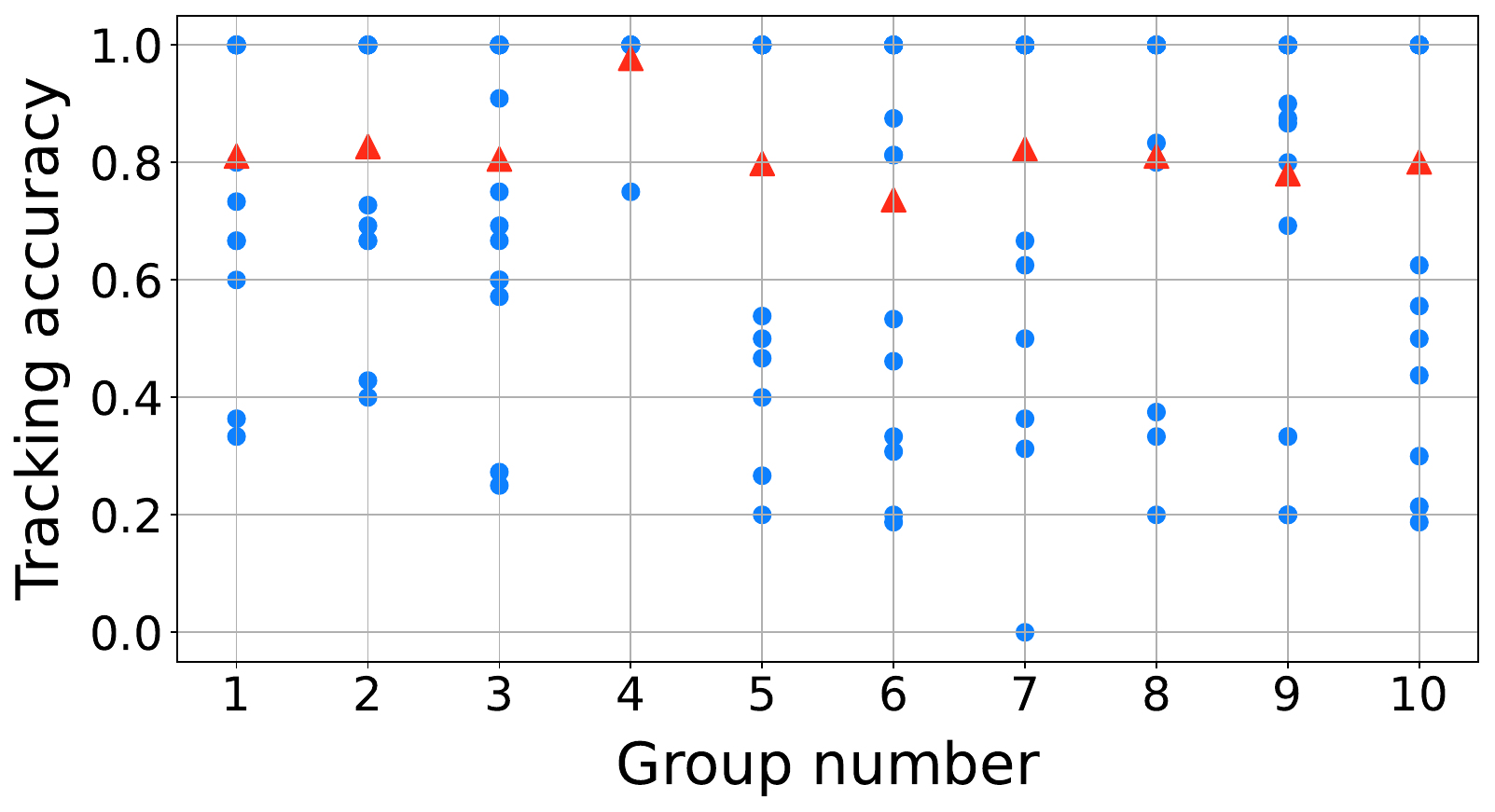}
    \caption{
    The performance of our tracking method. Each blue dot represents a single filament tracking accuracy. The red triangles represent the average accuracy of each group.}
    \label{fig9}
\end{figure}

\begin{table*}[!ht]
    \centering
    \caption{Performance of our tracking method.}    
    \label{tab3}
    \begin{tabular}{c|c|c|c|c|c|c|c|c|c|c}
    \hline
        Group number & 1& 2& 3& 4& 5& 6& 7& 8& 9& 10  \\ 
        \hline
        Filament numbers &15&14&19& 12& 18& 17& 20& 13& 22& 21  \\ 
        \hline
        Average accuracy &0.81 &0.83 &0.81 &0.98 &0.80 &0.74 &0.82 &0.81 &0.78 &0.80  \\ 
        \hline
    \end{tabular}
\end{table*}

We selected 10 groups of CHASE observation for testing the tracking performance. Each group is the first day of a month and has about 15 frames of H$\alpha$ line center images with time interval about one hour. If the filament are tracked by our tracking method in more than 3 frames, it will be counted for the tracking accuracy testing. If the tracking ID changed but the ground-truth is the same filament, it would be regarded as false tracking. The testing results are plotted in Figure~\ref{fig9} and summarized in Table~\ref{tab3}. The average tracking accuracy is $81.7\%$, which confirm the good performance of our tracking method. 

\section{discussion and conclusion} \label{discussion&conclusion}

Based on the characteristics of CHASE observations, we have developed an efficient automated method for detecting, tracking and analyzing filaments\footnote{The code of our detection, tracking, and analysis methods are available on GitHub (\url{https://github.com/ZZsolar/filament-detection-and-tracking.git} and the data are available in Solar Science Data Center of Nanjing University  (\url{https://ssdc.nju.edu.cn/NdchaseSatellite}). The code and data have also been deposited to Zenodo under a Creative Commons Attribution license: \dataset[doi:10.5281/zenodo.10598419]{https://doi.org/10.5281/zenodo.10598419}.}. Instead of manually annotating filaments, we use the K-Means method as our spectral classification tool combined with morphological open operation to obtain the labeled filament, which guarantees the consistence and accuracy of labelling. Then these labeled data are adopt to train the U-Net model, and the good performance demonstrates that our method is a viable strategy for solar filament detection. It is note that the K-means method can also be adopted for filament detection. We adopt U-net model for filament detection instead of K-means method due to two main reasons. First, the K-means method is very sensitive to the uneven radiation intensity distribution across the solar disk, which is adversely affecting spectral classification for labeling filaments. It means that we have to remove the limb darkening firstly. Further more,  for each file, we should select the class of the filament manually, i.e., the K-means algorithm need additional manual assistant. So if we use K-means method for detecting, our whole detection module would be semi-automated. Second, compared to using H$\alpha$ spectral data as the input of K-means method during the preprocessing steps, we only adopt the original H$\alpha$ line center image as the input of U-Net model. In our test, the data processing and spectral classification of a H$\alpha$ spectral file by K-means method takes several minutes, while the detection by the U-Net model takes less than one second. It greatly minimizes the time consumption of filament detection.

After the filament recognition, besides getting the ordinary morphological features such as filament area and location, we apply the graph theory to extract the filament spine, and use the cloud model to inverse the LOS velocity distribution of the filament region. In a follow-up work, we will implement our filament feature extraction and analysis methods to do a statistical study on filament features, not only on their morphological features, but also on the dynamic evolution, which are valuable for a better understanding of the physical mechanisms of filaments. We also integrate the U-Net model and the CSRT algorithm to track the evolution of filaments, the test results show that our tracking method can track filaments efficiently regardless of whether they are experiencing motions, deformations, breaking, and eruptions.

Although our method is performed well, we also found some limitations in our work during the experiment. First, sometimes the K-Means spectral classification cannot figure out relative small filaments which are surrounded by plages, as indicated in Figure~\ref{fig2}(c) and (d). These filaments can not be separated using intensity thresholds in H$\alpha$ line center images. The reason may be that the K-Means algorithm only considers the distances between different classes, while it does not account for the weights of factors such as line depth and line width. Some other unsupervised pre-classification models would be considered as the pre-labeling method to take account of more information of H$\alpha$ spectra. Second, our detection method based on the U-Net architecture, which has an excellent performance in semantic segmentation, cannot handle the problem of filament fragments very well (if the fragments belong to one filament without connection with each other). This problem can affect the tracking accuracy since the filaments detected by the U-Net model are used as inputs of the CSRT tracker. For example, if the first frame misidentifies the filament with several fragments, the following tracking procedure can only treat the fragments as separate filaments. The possible ways to solve this problem require to add more information in the labeled data and one more neural networks to merge the fragments belonging to one filament, or adopt the instance-segmentation model like that of \citet{Guo2022} to solve the fragment problems. 

In summary, we have developed an efficient automated method for filament detection and tracking. We utilized the U-Net model to identify filaments and implemented the Channel and Spatial Reliability Tracking (CSRT) algorithm for automated filament tracking. In addition, we used the cloud model to invert the LOS velocity of filament and employed the graph theory algorithm to extract the filament spine, which can promote understanding the dynamics of filaments. The test favorable performance confirms the validity of our method, which will be implemented in the following statistical analyses of filament features and dynamics based on CHASE observations.

\begin{acknowledgements}
CHASE mission was supported by China National Space Administration. This work was supported by NSFC under grants 12173019, 12333009, 12127901, and the CNSA project D050101, as well as the AI \& AI for Science Project of Nanjing University.
\end{acknowledgements}

\bibliography{filamentMS}{}

\begin{thebibliography}{}
\expandafter\ifx\csname natexlab\endcsname\relax\def\natexlab#1{#1}\fi
\providecommand{\url}[1]{\href{#1}{#1}}
\providecommand{\dodoi}[1]{doi:~\href{http://doi.org/#1}{\nolinkurl{#1}}}
\providecommand{\doeprint}[1]{\href{http://ascl.net/#1}{\nolinkurl{http://ascl.net/#1}}}
\providecommand{\doarXiv}[1]{\href{https://arxiv.org/abs/#1}{\nolinkurl{https://arxiv.org/abs/#1}}}

\bibitem[{{Asensio Ramos} {et~al.}(2023){Asensio Ramos}, {Cheung}, {Chifu}, \&
  {Gafeira}}]{Asensio2023}
{Asensio Ramos}, A., {Cheung}, M. C.~M., {Chifu}, I., \& {Gafeira}, R. 2023,
  Living Reviews in Solar Physics, 20, 4, \dodoi{10.1007/s41116-023-00038-x}

\bibitem[{{Beckers}(1964)}]{Beckers1964}
{Beckers}, J.~M. 1964, PhD thesis, National Solar Observatory, Sunspot New
  Mexico

\bibitem[{{Bernasconi} {et~al.}(2005){Bernasconi}, {Rust}, \&
  {Hakim}}]{Bernasconi2005}
{Bernasconi}, P.~N., {Rust}, D.~M., \& {Hakim}, D. 2005, Solar Phys, 228, 97,
  \dodoi{10.1007/s11207-005-2766-y}

\bibitem[{{Bonnin} {et~al.}(2013){Bonnin}, {Aboudarham}, {Fuller}, {Csillaghy},
  \& {Bentley}}]{Bonnin2013}
{Bonnin}, X., {Aboudarham}, J., {Fuller}, N., {Csillaghy}, A., \& {Bentley}, R.
  2013, \solphys, 283, 49, \dodoi{10.1007/s11207-012-9985-9}

\bibitem[{Bradski(2000)}]{opencv}
Bradski, G. 2000, Dr. Dobb's Journal of Software Tools

\bibitem[{{Chatzistergos, Theodosios} {et~al.}(2023){Chatzistergos,
  Theodosios}, {Ermolli, Ilaria}, {Banerjee, Dipankar}, {Barata, Teresa},
  {Chouinavas, Ioannis}, {Falco, Mariachiara}, {Gafeira, Ricardo}, {Giorgi,
  Fabrizio}, {Hanaoka, Yoichiro}, {Krivova, Natalie A.}, {Korokhin, Viktor V.},
  {Lourenço, Ana}, {Marchenko, Gennady P.}, {Malherbe, Jean-Marie}, {Peixinho,
  Nuno}, {Romano, Paolo}, \& {Sakurai, Takashi}}]{Chatzistergos2023}
{Chatzistergos, Theodosios}, {Ermolli, Ilaria}, {Banerjee, Dipankar}, {et~al.}
  2023, \aap, 680, A15, \dodoi{10.1051/0004-6361/202347536}

\bibitem[{{Chen}(2011)}]{Chen2011}
{Chen}, P.~F. 2011, Living Reviews in Solar Physics, 8, 1,
  \dodoi{10.12942/lrsp-2011-1}

\bibitem[{{Chen} {et~al.}(2020){Chen}, {Xu}, \& {Ding}}]{Chen2020}
{Chen}, P.-F., {Xu}, A.-A., \& {Ding}, M.-D. 2020, Research in Astronomy and
  Astrophysics, 20, 166, \dodoi{10.1088/1674-4527/20/10/166}

\bibitem[{Farhodov {et~al.}(2019)Farhodov, Kwon, Kang, Lee, \&
  Kwon}]{Farhodov2019}
Farhodov, X., Kwon, O.-H., Kang, K.~W., Lee, S.-H., \& Kwon, K.-R. 2019, in
  2019 International Conference on Information Science and Communications
  Technologies (ICISCT), 1--3, \dodoi{10.1109/ICISCT47635.2019.9012043}

\bibitem[{{Fuller} {et~al.}(2005){Fuller}, {Aboudarham}, \&
  {Bentley}}]{Fuller2005}
{Fuller}, N., {Aboudarham}, J., \& {Bentley}, R.~D. 2005, Solar Phys, 227, 61,
  \dodoi{10.1007/s11207-005-8364-1}

\bibitem[{{Gao} {et~al.}(2002){Gao}, {Wang}, \& {Zhou}}]{Gao2002}
{Gao}, J., {Wang}, H., \& {Zhou}, M. 2002, Solar Phys, 205, 93

\bibitem[{{Gopalswamy} {et~al.}(2003){Gopalswamy}, {Shimojo}, {Lu}, {Yashiro},
  {Shibasaki}, \& {Howard}}]{Gopalswamy2003}
{Gopalswamy}, N., {Shimojo}, M., {Lu}, W., {et~al.} 2003, \apj, 586, 562,
  \dodoi{10.1086/367614}

\bibitem[{{Guo} {et~al.}(2022){Guo}, {Yang}, {Feng}, {Bai}, {Liang}, \&
  {Dai}}]{Guo2022}
{Guo}, X., {Yang}, Y., {Feng}, S., {et~al.} 2022, \solphys, 297, 104,
  \dodoi{10.1007/s11207-022-02019-z}

\bibitem[{{Hao} {et~al.}(2015){Hao}, {Fang}, {Cao}, \& {Chen}}]{Hao2015}
{Hao}, Q., {Fang}, C., {Cao}, W., \& {Chen}, P.~F. 2015, \apjs, 221, 33,
  \dodoi{10.1088/0067-0049/221/2/33}

\bibitem[{{Hao} {et~al.}(2013){Hao}, {Fang}, \& {Chen}}]{Hao2013}
{Hao}, Q., {Fang}, C., \& {Chen}, P.~F. 2013, \solphys, 286, 385,
  \dodoi{10.1007/s11207-013-0285-9}

\bibitem[{{Hao} {et~al.}(2016){Hao}, {Guo}, {Fang}, {Chen}, \& {Cao}}]{Hao2016}
{Hao}, Q., {Guo}, Y., {Fang}, C., {Chen}, P.-F., \& {Cao}, W.-D. 2016, Research
  in Astronomy and Astrophysics, 16, 1, \dodoi{10.1088/1674-4527/16/1/001}

\bibitem[{{Hou} {et~al.}(2023){Hou}, {Li}, {Li}, {Su}, {Qiu}, {Yang}, {Yang},
  {Li}, {Guo}, {Hou}, {Song}, {Bai}, {Zhou}, {Ding}, {Gan}, \&
  {Deng}}]{Hou2023}
{Hou}, Y., {Li}, C., {Li}, T., {et~al.} 2023, \apj, 959, 69,
  \dodoi{10.3847/1538-4357/ad08bd}

\bibitem[{{Kingma} \& {Ba}(2014)}]{Kingma2014}
{Kingma}, D.~P., \& {Ba}, J. 2014, arXiv e-prints, arXiv:1412.6980,
  \dodoi{10.48550/arXiv.1412.6980}

\bibitem[{Krizhevsky {et~al.}(2012)Krizhevsky, Sutskever, \&
  Hinton}]{krizhevsky2012}
Krizhevsky, A., Sutskever, I., \& Hinton, G.~E. 2012, Advances in neural
  information processing systems, 25, 1097

\bibitem[{{Labrosse} {et~al.}(2010){Labrosse}, {Heinzel}, {Vial}, {Kucera},
  {Parenti}, {Gun{\'a}r}, {Schmieder}, \& {Kilper}}]{Labrosse2010}
{Labrosse}, N., {Heinzel}, P., {Vial}, J.~C., {et~al.} 2010, \ssr, 151, 243,
  \dodoi{10.1007/s11214-010-9630-6}

\bibitem[{{Li} {et~al.}(2019){Li}, {Fang}, {Li}, {Ding}, {Chen}, {Chen}, {Lin},
  {Chen}, {Chen}, {Tao}, {You}, {Hao}, {Dai}, {Cheng}, {Guo}, {Hong}, {An},
  {Cheng}, {Chen}, {Wang}, \& {Zhang}}]{Li2019}
{Li}, C., {Fang}, C., {Li}, Z., {et~al.} 2019, Research in Astronomy and
  Astrophysics, 19, 165, \dodoi{10.1088/1674-4527/19/11/165}

\bibitem[{{Li} {et~al.}(2022){Li}, {Fang}, {Li}, {Ding}, {Chen}, {Qiu}, {You},
  {Yuan}, {An}, {Tao}, {Li}, {Chen}, {Liu}, {Mei}, {Yang}, {Zhang}, {Cheng},
  {Chen}, {Chen}, {Gu}, {Huang}, {Liu}, {Han}, {Xin}, {Chen}, {Ni}, {Wang},
  {Rao}, {Li}, {Lu}, {Wang}, {Lin}, {Jiang}, {Meng}, \& {Zhao}}]{Li2022}
---. 2022, Science China Physics, Mechanics, and Astronomy, 65, 289602,
  \dodoi{10.1007/s11433-022-1893-3}

\bibitem[{Lin {et~al.}(2017)Lin, Goyal, Girshick, He, \& Dollár}]{Lin2017}
Lin, T.-Y., Goyal, P., Girshick, R., He, K., \& Dollár, P. 2017, in 2017 IEEE
  International Conference on Computer Vision (ICCV), 2999--3007,
  \dodoi{10.1109/ICCV.2017.324}

\bibitem[{{Liu} {et~al.}(2021){Liu}, {Song}, {Lin}, \& {Wang}}]{Liu2021}
{Liu}, D., {Song}, W., {Lin}, G., \& {Wang}, H. 2021, \solphys, 296, 176,
  \dodoi{10.1007/s11207-021-01920-3}

\bibitem[{{Liu} {et~al.}(2022){Liu}, {Tao}, {Chen}, {Han}, {Chen}, {Mei},
  {Yang}, {Hu}, {Xin}, {Li}, {Guan}, {Xue}, {Zhu}, {Hu}, {Ha}, {He}, {Fang},
  {Li}, \& {Li}}]{Liu2022}
{Liu}, Q., {Tao}, H., {Chen}, C., {et~al.} 2022, Science China Physics,
  Mechanics, and Astronomy, 65, 289605, \dodoi{10.1007/s11433-022-1917-1}

\bibitem[{{Liu} {et~al.}(2012){Liu}, {Kliem}, {T{\"o}r{\"o}k}, {Liu}, {Titov},
  {Lionello}, {Linker}, \& {Wang}}]{Liu2012}
{Liu}, R., {Kliem}, B., {T{\"o}r{\"o}k}, T., {et~al.} 2012, \apj, 756, 59,
  \dodoi{10.1088/0004-637X/756/1/59}

\bibitem[{Lukežic {et~al.}(2017)Lukežic, Vojír, Zajc, Matas, \&
  Kristan}]{Lukezic2017}
Lukežic, A., Vojír, T., Zajc, L.~C., Matas, J., \& Kristan, M. 2017, in 2017
  IEEE Conference on Computer Vision and Pattern Recognition (CVPR),
  4847--4856, \dodoi{10.1109/CVPR.2017.515}

\bibitem[{MacQueen(1967)}]{MacQueen1967}
MacQueen, J. 1967, in Proceedings of the Fifth Berkeley Symposium on
  Mathematical Statistics and Probability, Volume I-Theory of Statistics
  (University of California Press), 281–297.
\newblock
  \url{https://digitalassets.lib.berkeley.edu/math/ucb/text/math_s5_v1_frontmatter.pdf}

\bibitem[{{Martin}(1998)}]{Martin1998}
{Martin}, S.~F. 1998, \solphys, 182, 107, \dodoi{10.1023/A:1005026814076}

\bibitem[{Mikołajczyk \& Grochowski(2018)}]{Miko2018}
Mikołajczyk, A., \& Grochowski, M. 2018, in 2018 International
  Interdisciplinary PhD Workshop (IIPhDW), 117--122,
  \dodoi{10.1109/IIPHDW.2018.8388338}

\bibitem[{{Panos} {et~al.}(2018){Panos}, {Kleint}, {Huwyler}, {Krucker},
  {Melchior}, {Ullmann}, \& {Voloshynovskiy}}]{Panos2018}
{Panos}, B., {Kleint}, L., {Huwyler}, C., {et~al.} 2018, \apj, 861, 62,
  \dodoi{10.3847/1538-4357/aac779}

\bibitem[{{Pierce} \& {Slaughter}(1977)}]{Pierce1977}
{Pierce}, A.~K., \& {Slaughter}, C.~D. 1977, \solphys, 51, 25,
  \dodoi{10.1007/BF00240442}

\bibitem[{{Qiu} {et~al.}(2022){Qiu}, {Rao}, {Li}, {Fang}, {Ding}, {Li}, {Ni},
  {Wang}, {Hong}, {Hao}, {Dai}, {Chen}, {Wan}, {Xu}, {You}, {Yuan}, {Tao},
  {Li}, {He}, \& {Liu}}]{Qiu2022}
{Qiu}, Y., {Rao}, S., {Li}, C., {et~al.} 2022, Science China Physics,
  Mechanics, and Astronomy, 65, 289603, \dodoi{10.1007/s11433-022-1900-5}

\bibitem[{{Qu} {et~al.}(2005){Qu}, {Shih}, {Jing}, \& {Wang}}]{Qu2005}
{Qu}, M., {Shih}, F.~Y., {Jing}, J., \& {Wang}, H. 2005, Solar Phys, 228, 119,
  \dodoi{10.1007/s11207-005-5780-1}

\bibitem[{Rezatofighi {et~al.}(2019)Rezatofighi, Tsoi, Gwak, Sadeghian, Reid,
  \& Savarese}]{Rezatofighi2019}
Rezatofighi, H., Tsoi, N., Gwak, J., {et~al.} 2019, Generalized Intersection
  over Union: A Metric and A Loss for Bounding Box Regression.
\newblock \doarXiv{1902.09630}

\bibitem[{{Ronneberger} {et~al.}(2015){Ronneberger}, {Fischer}, \&
  {Brox}}]{Ronneberger2015}
{Ronneberger}, O., {Fischer}, P., \& {Brox}, T. 2015, arXiv e-prints,
  arXiv:1505.04597, \dodoi{10.48550/arXiv.1505.04597}

\bibitem[{{Shen} {et~al.}(2012){Shen}, {Liu}, \& {Su}}]{Shen2012}
{Shen}, Y., {Liu}, Y., \& {Su}, J. 2012, \apj, 750, 12,
  \dodoi{10.1088/0004-637X/750/1/12}

\bibitem[{{Shih} \& {Kowalski}(2003)}]{Shih2003}
{Shih}, F.~Y., \& {Kowalski}, A.~J. 2003, \solphys, 218, 99,
  \dodoi{10.1023/B:SOLA.0000013052.34180.58}

\bibitem[{{Smith} \& {Geach}(2023)}]{Smith2023}
{Smith}, M.~J., \& {Geach}, J.~E. 2023, Royal Society Open Science, 10, 221454,
  \dodoi{10.1098/rsos.221454}

\bibitem[{{Sun} {et~al.}(2023){Sun}, {Li}, {Tian}, {Hou}, {Hou}, {Chen}, {Bai},
  \& {Deng}}]{Sun2023}
{Sun}, Z., {Li}, T., {Tian}, H., {et~al.} 2023, \apj, 953, 148,
  \dodoi{10.3847/1538-4357/ace5b1}

\bibitem[{Tian {et~al.}(2020)Tian, Shen, \& Chen}]{Tian2020}
Tian, Z., Shen, C., \& Chen, H. 2020, in Computer Vision -- ECCV 2020, ed.
  A.~Vedaldi, H.~Bischof, T.~Brox, \& J.-M. Frahm (Cham: Springer International
  Publishing), 282--298

\bibitem[{Tian {et~al.}(2023)Tian, Zhang, Chen, \& Shen}]{Tian2023}
Tian, Z., Zhang, B., Chen, H., \& Shen, C. 2023, IEEE Transactions on Pattern
  Analysis and Machine Intelligence, 45, 669,
  \dodoi{10.1109/TPAMI.2022.3145407}

\bibitem[{{Vial} \& {Engvold}(2015)}]{Vial2015}
{Vial}, J.~C., \& {Engvold}, O., eds. 2015, Astrophysics and Space Science
  Library, Vol. 415, {Solar Prominences}, \dodoi{10.1007/978-3-319-10416-4}

\bibitem[{{Viticchi{\'e}} \& {S{\'a}nchez Almeida}(2011)}]{Viticchie2011}
{Viticchi{\'e}}, B., \& {S{\'a}nchez Almeida}, J. 2011, \aap, 530, A14,
  \dodoi{10.1051/0004-6361/201016096}

\bibitem[{{Wang} {et~al.}(2010){Wang}, {Cao}, {Chen}, {Zhang}, {Yu}, {Zheng},
  {Shen}, {Zhang}, \& {Wang}}]{Wang2010}
{Wang}, Y., {Cao}, H., {Chen}, J., {et~al.} 2010, \apj, 717, 973,
  \dodoi{10.1088/0004-637X/717/2/973}

\bibitem[{{Yu} {et~al.}(2020){Yu}, {Li}, {Ding}, {Li}, {Zhou}, \&
  {Hong}}]{Yu2020}
{Yu}, K., {Li}, Y., {Ding}, M.~D., {et~al.} 2020, \apj, 896, 154,
  \dodoi{10.3847/1538-4357/ab9014}

\bibitem[{{Yuan} {et~al.}(2011){Yuan}, {Shih}, {Jing}, {Wang}, \&
  {Chae}}]{Yuan2011}
{Yuan}, Y., {Shih}, F.~Y., {Jing}, J., {Wang}, H., \& {Chae}, J. 2011, Solar
  Phys, 272, 101, \dodoi{10.1007/s11207-011-9798-2}

\bibitem[{{Zhang} {et~al.}(2023){Zhang}, {Hao}, \& {Chen}}]{Zhang2023}
{Zhang}, T., {Hao}, Q., \& {Chen}, P.~F. 2023, \apjs

\bibitem[{{Zhu} {et~al.}(2019){Zhu}, {Lin}, {Wang}, {Liu}, \& {Yang}}]{Zhu2019}
{Zhu}, G., {Lin}, G., {Wang}, D., {Liu}, S., \& {Yang}, X. 2019, \solphys, 294,
  117, \dodoi{10.1007/s11207-019-1517-4}

\end{thebibliography}
\bibliographystyle{aasjournal}

\end{document}